# Research on innovation in China and Latin America: Bibliometric insights in the field of business, management and decision sciences

Julián David Cortés-Sánchez [a,b]*

[a]*School of Management and Business, Universidad del Rosario, Bogotá, Colombia*

[b]*Fudan Development Institute, Fudan University, Shanghai, China*

*Corresponding author email: julian.cortess@urosario.edu.co

Corresponding author address: Calle 200 con Autopista Norte, Bogotá, Colombia

- Julián David Cortés-Sánchez is a principal professor at Universidad del Rosario's School of Management and Business (Colombia), and visiting scholar (2019) and invited researcher at Fudan University's Development Institute (China).

# Research on innovation in China and Latin America: Bibliometric insights in the field of business, management and decision sciences

China and Latin America (LATAM) are now key players in global research production. This study presents a comparative study the on research on innovation in management and decision sciences based on data from Scopus and Web of Knowledge (WoS) between China and LATAM. Findings showed significant differences between regions regarding journals' citation dependent measures, and between the number of authors and journal reputation, public universities have been leading producers, and China showed a particular interest in research topics such as *commerce* and *industry*, while LATAM in *sustainable development* and *bio-technology.*

Keywords: bibliometrics; innovation; business; management; China; Latin America.

## 1 Introduction and research background

Commercial and knowledge exchange within the Global South (i.e., countries located in Asia, Africa, Latin America and the Caribbean [LATAM]) is becoming increasingly relevant (Ortiz-Ospina, Beltekian, & Roser, 2018). A sustainable commercial and knowledge exchange agenda in the Global South considers the promotion of inclusive and sustainable industrialization fostered by innovation (UN, 2018). In the context of studies related to management and business, innovation can be understood as: "the invention and implementation of a management practice, process, structure, or technique that is new to the state of the art and is intended to further organizational goals" (Birkinshaw et al. 2008 p. 825). Despite the setbacks in certain indicators (e.g., developing regions falling short of the 1.68% GDP world average in R&D investment (United Nations, 2019)), research output related to innovation in business and management (iBM) and decision sciences (DS) has been consistently increasing in the Global South (Tollefson, 2018). Research on iBM in both China and LATAM has reached the number of +1,300 and +1,000 articles from 1996 to 2018 (Scopus, 2018), respectively. That substantial research substrate gains relevance as the appropriation of research on iBM is related to firms' growth (Rosenbusch, Brinckmann, & Bausch, 2011).

To both researchers and practitioners, several concerns may arise by counting the number of research entities for the thousands, such as identifying the trending research topics and the structure of both the innovation-related fields and the researcher/institutional social capital? Are there consolidated research topics? Which are the (new) research fronts? Which are the key authors/institutions in the social capital network to build strategic alliances with? Methods from bibliometrics such as citation analysis, keywords co-occurrence, and co-citation networks, allow answering those inquiries (Zupic & Čater, 2015).

To settle the research background, two streams of literature in bibliometrics/scientometrics were reviewed: comparative studies on iBM-DS, and studies focused on China and LATAM. Regarding the first stream, studies on nanoscience/technology have found a consistent increase in China's research output, which was doubling every 2.1 years, surpassing the US in 2012 (Kostoff, 2012) but maintaining a low level of citations (Guan & Ma, 2007). Tijssen (2009) argued that research from European pharmaceuticals was oriented towards the US science base than vice versa. When analyzing two biotechnology clusters in UK and Germany, Casper and Murray (2005) found that the networks of scientists from the UK contained a balanced mix of industry and scientific experience. German scientists' experience, however, was more academy-based. An analysis of 21 disciplines in 34 countries and their revealed comparative advantage (RCA) conducted by Harzing and Giroud (2014) proposed the clustering of five groups: G1 (most populous English-speaking countries): social sciences; G2, 5 (France, Italy, Poland, Russia and Ukraine): physical sciences; G3 (Germany, Austria and Switzerland): balanced research profile; G4 (Asian Tigers, except for Hong-Kong and China): engineering. Subsequent studies suggested for the Global South to choose and specialized in specific scientific fields to increase research impact (Confraria, Mira Godinho, & Wang, 2017; de Paulo, Carvalho, Costa, Lopes, & Galina, 2017). Concretely in the field of technology and innovation management (TIM), studies have identified the intellectual pillars in the field (e.g., dynamic organizations, innovation process, knowledge management) (Pilkington & Teichert, 2006) and specific research interest depending on the regions, such as *organization; technology strategy; new product development, design and innovation; technology policy;* and *technological acquisition* in developed countries, whereas *technology policy; organization; technological acquisition; R&D management; technological change* and *technological development* were the most researched by authors from developing countries (Cetindamar, Wasti, Ansal, & Beyhan, 2009). Choi et al. (2012), with an aligned aim, argued that among European countries, the UK has a comparative advantage in *social change*; Spain in *intellectual property*; the Netherlands in *technology policy*; Germany in *entrepreneurship*; and Italy in *technology transfer and commercialization*.

Regarding the second stream of studies focused on China and LATAM, when studying the relationship between science and technology in China, Guan and He (2007) pointed that patents from sectors such as biotech, pharmaceutics and organic chemistry cite more scientific papers. In contrast, patents from sectors such as information and communication technologies, semi-conductors and optics cite other patents more often. Subsequent analyses were focused on the inclusion of technology foresight and road-mapping in national and regional policy design-applications in science, technology and innovation (N. Li, Chen, & Kou, 2017), and recommendations for supporting emerging technologies (e.g., solar cell industry) (L. Huang, Zhang, Guo, Zhu, & Porter, 2014; X. Li, Zhou, Xue, & Huang, 2015). In the case of LATAM, research has followed a diversified agenda, including Schumpeterian innovation and cooperation (Lazzarotti, Dalfovo, & Hoffmann, 2011; Lopes & De Carvalho, 2012), innovativeness measures (De Carvalho, Cruz, De Carvalho, Duclós, & De Fátima Stankowitz, 2017), industry relations (Manjarrez, Pico, & Díaz, 2016), business models (Ceretta, Dos Reis, & Da Rocha, 2016), financing on innovation (Padilla-Ospina, Medina-Vásquez, &

Rivera-Godoy, 2018), social innovation (Silveira & Zilber, 2017), supply chain management (Tanco, Escuder, Heckmann, Jurburg, & Velazquez, 2018) and the citation, production, and academic-corporate collaboration (Cortés-Sánchez, 2019). Highlighted findings stated worldwide maturity in industrial relations with innovation system players (e.g., academic, scientific or technological) not so in LATAM, except for Brazil; five salient topics in financing innovation (i.e., financial constraints, funding sources, capital structure, venture capital, and financing of technology companies); the research output on supply chain management in LATAM has not been substantial for the field from a global perspective in terms of output, citations, presence in top-tier journals, and both international and corporate collaboration. Further insights showed an intensive use of a journal with predatory features, henceforth citations/documents measures are in the lowest of the last decade, documents which leading author was affiliated with non-LATAM institutions were significantly more cited in average, and despite the barely noticeable academic-corporate collaboration, multinationals and central-national organism were involved (e.g., Central Banks).

In sum, bibliometric studies in China and LATAM in the fields of iBM-DS have been analyzing intensive-knowledge industries, the global north-south relationship dynamics, technology and innovation management, innovativeness measures, and social and open innovation. That research stream entails two similarities: the relationship dynamics Global North-South and the use of Web of Science (WoS) as the pervasive data source for peer-reviewed references and citations. Despite the findings presented above, studies analyzing the relation within the Global South using complementary data to WoS have not been investigated with the same intensity. Therefore, this study seeks to deepen the understanding of the research dynamics on iBM-DS published by authors affiliated with institutions from China and LATAM in a comparative framework using both WoS and Scopus. The rest of this paper is organized as follows: The methodology is presented in the next section. Afterward, the results are both analyzed and discussed. Finally, the conclusions are outlined.

## 2 Methodology

### 2.1 Data

Data was gathered from two different sources: WoS and Scopus. The former had been the most used source for bibliometrics and scientometrics research historically, while the spotlight is getting wider for the latter in the last 15 years (Archambault, Campbell, Gingras, & Larivière, 2009). WoS possesses over 100 million records from 33,000 journals (Clarivate Analytics, 2017). For its part, Scopus covers over 75 million items (i.e., articles, proceedings, books) published by over 5,000 publishers, authored by 16 million authors affiliated to more than 70,000 institutions (Scopus, 2019). Comparisons between both systems conclude that they provide robust and accurate data for the items covered (Amara & Landry, 2012; Mingers & Lipitakis, 2010) with the caveat that Scopus has a wider journal coverage (i.e., Scopus: 20,346 journals vs. WoS: 13,605) of both articles and journals published by countries in Ibero-America (e.g., LATAM, Spain and Portugal) and a greater social sciences coverage (i.e., Scopus: ≈25% vs. WoS: ≈15% ) (Gavel & Iselid, 2008; Mongeon & Paul-Hus, 2016). Table 1 summarizes the search query used in each bibliographic database and the results. Over 300 articles of the journal *Espacios* were excluded since this journal unveils predatory-journal features (Beall, 2015; Cortés-Sánchez, 2018). The complete dataset is available at the end of this article as a permanent link.

[Table 1]

### 2.2 Methods

Research output, citations, publishing market overview and one-way ANOVAs (Analysis of Variance) to explore significant differences among regions, authors' social capital, journals' h

Index and citations will be presented. IBM's SPSS V27 was used for the analysis. Co-occurrence and co-citation networks were generated by VOSviewer (van Eck & Waltman, 2010). Co-authorship networks are tools considered often in bibliometric studies. In this study, it was replaced by a closer analysis regarding the number of authors and significant differences in journals' citations related-measurements (i.e., Journal Impact Factor, Eigenvector, H Index).

## 3 Results

### 3.1 Research output and citations

Table 2 presents descriptive statistics by region (i.e., China and LATAM) considering WoS (i.e., articles' number of authors, Journals' Impact Factor and Eigenfactor, open access, and citations) and Scopus (i.e., articles' number of authors, journals' H index, open access and citations) variables. The Journal Impact Factor results from dividing the number of citations in the current year of a journal in the previous two years by the number of articles published by the same journal in the same two years (Garfield, 2015). The Eigenfactor score is a measure of a journals' importance in the scientific community (Bergstrom, West, & Wiseman, 2008). The Eigenfactor score interpretation is that if a journal has an Eigenfactor of 1.0, it means it has 1% of the total influence (Eigenfactor.org, 2019). The H index shows that if a journal has an H index of 20, if 20 of its articles have at least 20 citations, whereas the rest of the articles have less than 20 (Hirsch, 2005).

[Table 2]

Table 3 presents the ANOVAs summary and post hoc Tukey HSD tests, if significant differences between groups' means are detected, comparing both regions for the following variables for WoS: number of authors; citations; Journals' Impact Factor; and Eigenfactor. For Scopus, the variables considered were number of authors; citations; and journals' H

Index. Results for WoS showed no significant differences between regions for number of authors, there were, however, significant differences for citations, Journals' Impact Factor, and Eigenfactor. China showed a higher mean for the three latter variables. Results for Scopus also showed no significant differences for number of authors, there were, significant differences for journals' H Index and citations. Again, China showed a higher mean for the two latter variables. When authors were divided into four groups: individual, couples, trios, and packs (i.e., four or more authors) as proxies of the social capital of authors (Wuchty, Jones, & Uzzi, 2007), there were not found significant differences in WoS for citations nor Eigenfactor, yet there were significant differences for the Journal Impact Factor. The mean Journal Impact Factor was higher for the pack than the individual and the couple groups, and for the trios than the couples' groups. In addition, there was a positive correlation between the number of authors and Journal Impact Factor ($r=.049$, $p=.003$). However, there was no correlation between the number of authors and citations ($r=-.023$, $p=.175$), and Eigenfactor ($r=-.001$, $p=.953$). For Scopus, also there were no significant differences of groups of authors for citations but there were significant differences of group of authors for journals' H Index. The mean journals' H Index for the pack group was higher than the rest of the groups. There was a positive albeit small correlation between number of authors and Journals' H Index ($r=.109$, $p=.000$). There was no correlation between the number of authors and citations ($r=.022$, $p=.290$).

[Table 3]

Figure 1 presents the total output and China-LATAM ratio for both WoS and Scopus 2001-2018. For several years during 2001-2005 the production was of one digit for both regions. There is a clear increasing trend for both regions, where China has been the leading producer in almost every year for both WoS and Scopus. In Scopus, China reached hundreds of articles around 2014 and LATAM one year early. In WoS, China reached hundreds of articles around 2011 and LATAM until 2015. With that in mind, in comparative terms, there have been two peaks of production between China-LATAM in WoS. A closer look for 2005 and 2008 showed that China's production was nine and four times higher than the whole LATAM region, respectively. In the last five years, China has produced an average of 80% more articles than LATAM. There are subtle differences when looking at Scopus although. The highest production peak for China was 2004 with 2.5 the production of LATAM. However, LATAM outperformed China in 2003 (twice the production of China), 2009 (+20%), 2012 (+20%) and 2013 (+10%). For the last five years, China has produced an average of 50% more articles than LATAM. The most productive LATAM country was Brazil. A total of 802 articles were indexed in WoS have at least one (co)author with an affiliation of Brazil, 578 with the same features in Scopus. The top-five most productive countries in LATAM were Colombia (20% the production of Brazil in Scopus and 22% in WoS), Mexico (Scopus: 20%; WoS: 24%), Chile (Scopus: 14%; WoS: 15%) and Argentina (Scopus: 10%; WoS: 9%) (WoS, 2018; Scopus, 2018).

[Figure 1]

At the institutional level, public universities from LATAM and China are the leading institutions in production, with a few private exceptions (i.e., Tec de Monterrey and Fundação Getulio Vargas) (Figure 2). Most Chinese universities are part of the C9 League, an exceptional group of public universities (i.e., Zhejiang University [Scopus: 107; WoS:142], Xi'an Jiaotong University [Scopus: 61; WoS:112], Shanghai Jiao Tong University [Scopus: 57] and Tsinghua University [Scopus: 99; WoS:157]) (Australian Government - Department of Education, n.d.). Chinese institutions showed an inclination towards publishing in journals indexed in WoS. In LATAM, the most productive Brazilian universities in Scopus and WoS were Sao Paulo (135) and Estadual de Campinas (48).

**[Figure 2]**

Figure 3 presents the ratio citations/articles in periods of three years as in Bornmann et al. (2014). The average citation/articles showed a decreasing trend for both China and LATAM in WoS and Scopus. It is expected that mature articles concentrate the most citations. The period 2001-2003 showed the highest citation/articles ratio for China in both WoS (86.05) and Scopus (46.84). The trend for the subsequent periods followed a decreasing path. LATAM with a score of 47 surpassed China's score in Scopus (38.09) for the period 2004-2006.

**[Figure 3]**

### 3.2 *Highly cited publications*

Table 4 presents the top-ten (i.e., five from China and five from LATAM) most cited documents in each search engine/bibliographic database. A paper cited +400 times, authored by Stilgoe et al. (2013) and tracked by Scopus, was the most cited document published in LATAM. Most of the leading authors were affiliated with either a Chinese or European institution, followed by North-American and LATAM institutions. *Research Policy* published most of the highly cited articles listed, followed by the *Journal of Business Venturing, Journal of Marketing, Strategic Management Journal, Technovation,* and *World Development*. More than half of the articles listed were published between 2000 and 2007, mostly from 2004.

**[Table 4]**

### 3.3 *Co-occurrence and co-citations networks*

Co-occurrence and co-citations networks were generated exploiting each platform's characteristics. Co-occurrence networks were generated based on both Scopus and WoS; and a co-citation network based on WoS. A co-occurrence network displays the relatedness of items based on the number of documents in which they occur together (i.e., linking between keywords), generating a visualization of mature and emergent topics and their relatedness or divergence (van Eck & Waltman, 2010). The size of each node-keyword is weighted proportional to occurrences. A co-citation network connects academic disciplines by identifying at the reference section of a document and connecting the documents' research fields that appear in the references list (H. Small, Sweeney, & Greenlee, 1985). Among several other possibilities, co-citation networks allow understanding the development of scientific fields and the interrelationship among specialties (Small, 1973).

The co-occurrence network for China based on Scopus (Figure 4-a), consisted in 281 items, 6,460 links and seven clusters: red (83 items), green (65), blue (47), yellow (32), purple (26), pale blue (20), and orange (8). Keywords with high weight in each cluster were *technological development*; *industry, commerce, and competition; technological innovation; societies and institutions; decision making; product innovation; and investments*, respectively. The co-occurrence network for LATAM based on Scopus (Figure 4-b), consisted in 120 items, 1,649 links, and six clusters: red (39 items), green and blue (19), yellow (17), purple (15) and pale blue (11). Keywords with high weight in each cluster were *technological development*; *product development*; *competition*; *sustainable development*; *decision making;* and *biotechnology*. Similar keywords found in both China and LATAM were *technological development; product development/innovation; competition;* and *decision making*. China differentiated from LATAM in keywords such as *industry, commerce;* and *societies and institutions*. In LATAM, on the other hand, the differentiated keywords were: *sustainable development* and *biotechnology*.

**[Figure 4-a; Figure 4-b]**

The co-occurrence network for China based on WoS up to 500 references (Figure 5-a), consisted in 214 items, 5,070 links and seven clusters: red (48 items), green (42), blue (38), yellow (34), purple (29), pale blue (15), and orange (8). Keywords with high weight in each cluster were *firm performance* and *performance; knowledge; strategy; R&D; competitive advantage;* and *product innovation.* The co-occurrence network for LATAM (Figure 5-b), consisted in 194 items, 4,066 links, and six clusters: red (47 items), green (43), blue (37), yellow (31), purple (18), and pale blue (17). Keywords with high weight in each cluster were *knowledge; R&D; strategy and product development; dynamic capabilities; industry;* and *performance,* respectively. Both regions shared the following keywords: *performance; knowledge; R&D;* and *strategy.* Distinctive keywords for China were *product innovation;* and *competitive advantage*; and for LATAM: *product development; industry;* and *dynamic capabilities*.

The co-citation network of China (Figure 5-c) comprised of 203 items, 15,126 links, and four clusters: red (76 items), green (69), blue (49) and yellow (9). Journals with high weight in each cluster were: *Strategic Management Journal*; *Journal of Marketing*; *Journal of Management*; and *Academy of Management Review*. Co-citation network of LATAM (Figure 5-d) was composed of 186 items, 13,203 links, and six clusters: red (73 items), green (52), blue (21), yellow (18), purple (17), and blue pale (5). Journals with high weight in each cluster were: *Technovation*; *Research Policy*; *Entrepreneurship: Theory and Practice*; *Journal of Marketing*; *Academy of Management Journal*; and *Strategic Management Journal*. Both regions shared *Journal of Marketing* and *Strategic Management Journal* as central outlets for research.

**[Figure 5-a; Figure 5-b; Figure 5-c; Figure 5-d]**

## 4   Discussion

The mean number of authors and its increasing path is partially consistent with previous findings. In 2001, the average number of authors was 1.9 in Scopus and 2.2 in WoS, whereas in 2018 the average was 3.1 for both. This is a general trend since over the past 45 years the average number of authors per paper has increased from 1.9 to 3.5 (Wuchty et al., 2007). In management. Acedo et al. (2006) and Lazzarotti et al. (2011) calculated authors' mean in 2.8 and 1.8, respectively. That's partly because problems are more complex and need more specialized know-how (Kennedy, 2003). This finding is reinforced since the pack group showed significant differences for journals' Impact Factor and H Index. Findings might appear counterintuitive considering that the number of authors in China and LATAM (2013-2018) was +2,400 and +1,700, respectively. The number of authors may respond to a discipline's particularities, not to a net number of researchers available.

Significant differences between China and LATAM in citations and citations' dependent measures (i.e., Journals' Impact Factor, Eigenfactor and journals' H Index) are partly explained by previous citations, level of international collaboration and total publications in a specific scientific field (Confraria et al., 2017). Consider the latter factor, for instance. In this study, China outperformed LATAM in terms of output in both WoS and Scopus. On a broader discussion, China also surpassed the United States as the world leader in scientific production with 18.6% of Scopus's indexed studies in 2016 (Tollefson, 2018). However, current and future research quality in both regions should be scrutinized. The average citations/articles ratio of the last eight years was 69% below the average 2001-2009 ratio. In China, beyond the comparison with LATAM, the average number of citations for each article was 9.4, lower than the world average of 11.8 (F. Huang, 2018). As mentioned earlier, in the sample from LATAM, more than 300 articles were excluded because of predatory features.

At the institutional level, China's trend towards publishing in journals indexed in WoS may be partly explained by both pressure and financial incentives for publishing in the WoS Science Citation Index: "If a researcher does not publish at least half a dozen such articles and get national-level research funding as a principal investigator within their first five years as researchers, they have little hope of being hired as a tenured associate professor." Further, financial incentives for each publication are between US$900-10,000 (F. Huang, 2018).

The most cited papers and the co-citation analysis showed a considerable distance from the global-north incidence and the centrality and influence of certain journals. While the most cited papers of the region are far from the one thousand cites after more than a decade of being published, the three most cited papers on iBM in Scopus have +12,000 (Teece, Pisano, & Shuen, 1997) , +3.400 (Teece, 2007) and +3,200 (Leonard-Barton, 1992) cites, two were led by the same author (i.e, David Teece) who amassed +36,000 cites or 1.2 times the sum of the total citations of +2,000 articles analyzed in Scopus, and were authored by researchers affiliated to north American universities (i.e., Berkeley, Harvard, or San Jose State). The United States' foundational role in the creation of the first MBA programs and research fields (e.g., dynamic capabilities and strategic management) turned the country's intellectual formation and production into a global benchmark.

Co-citation results unveiled that both regions are orbiting around the subject categories of business and international management, marketing and strategy and management. With the caveat that China is also consulting advances from other research fields, such as psychology. The Journal of Applied Psychology was co-cited with other journals from the field of management (e.g., Journal of Management). Considering that the former journal: "*considers empirical and theoretical investigations that enhance understanding of cognitive, motivational, affective, and behavioral psychological phenomena*" (SCImago, 2018), such frameworks are enriching the phenomena studied in management and decision sciences (i.e., information systems, operation research and statistics). Co-citations from LATAM also considered studies from psychology and other subjects (i.e., economics and finance), however the node size is not as representative as in China. The most influential journals were published by a North American association (e.g., American Psychological Association;

Academy of Management; American Marketing Association) or one of the few publishers associated to the *oligopoly of academic publishers* (i.e., Reed-Elsevier, Wiley-Blackwell, Springer, and Taylor & Francis) (Larivière, Haustein, & Mongeon, 2015). Despite a substantial portion of research from LATAM was published in open access by public or private universities (e.g., Universidad Alberto Hurtado-Chile, Universidade Federal de Sao Carlos-Brazil, Universidad Nacional-Colombia: 26% of research from LATAM) journals from the region are not noticeable. The pioneering research on management and innovation in the Global North (Godin, 2015) and subsequent Matthew Effect can partly explain this (Merton, 1968).

Co-occurrence analysis showed both regional similitudes and differences. While key terms such as *technological development; product development/innovation;* and *competition* were relevant for both regions in Scopus, *performance;* and *knowledge;* were central in WoS. Both bibliographic databases shared key terms such as *decision making* and *strategy*, and *technological development* and *R&D*. This can amplify previous co-occurrence analysis focused just in LATAM as none of them were founded in previous studies (Cortés-Sánchez, 2019). China, on the one hand, showed a marked interest in key terms related to *industry* and *commerce,* which is linked to the country's priorities since the industry represents +40% of China's GDP (Statista, 2018) and is the largest export economy in the world (i.e., +2 trillion USD) (OEC, 2018). Management researchers from the region are studying *industry* focusing on green innovation drivers (Qi, Shen, Zeng, & Jorge, 2010) and *commerce* focusing on e-commerce (Peng & Lai, 2014). Previous text-mining analysis in LATAM (Favaretto & Francisco, 2017) confirmed consistency in research key terms (e.g., industry) and the emergence of several others such as *sustainable development* and *biotechnology*, both central for the regional context. LATAM is the most bio-diverse region in the world: 60% of the planet's life lives in the region (UNEP-WCMC, 2016). Researchers in management from the region are studying biodiversity under the frameworks of *sustainable development* focusing on cleaner production (Bonilla, Almeida, Giannetti, & Huisingh, 2010) or Corporate Social Responsibility (Anser, Zhang, & Kanwal, 2018), and *biotechnology* focusing on patents (Mendes, Amorim-Borher, & Lage, 2013) or university-industry interaction (Villasana, 2012), for instance.

## 5 Conclusion

The Global-South is gaining a place in global knowledge production. This knowledge's appropriation is crucial for the global agenda for development, particularly that related to iBM-DS. In this study, we contributed to both researcher and practitioners by presenting a comparative analysis on the research dynamics (i.e., research output, citations, authorship, highly cited studies, reputable journals, significant differences among regions, research topics relations and research fronts) on iBM-DS authored by authors affiliated with institutions from China and LATAM. The open access dataset may be used for further studies to replicate or triangulate the results or conduct analyses focused on a given country, in the case of LATAM, institutional, author level, or journal level.

The findings presented highlighted the relevance of increasing co-authorship for publishing in highly reputed journals. However, further impact could be explained by other factors, the historical and production dynamics for identifying gaps and how far are they, which are the causes of those gaps (e.g., the incentives for output oriented the easy-fast-low-quality publication in a predatory journal) and the key regional institutions/authors of the field, the journal-subject research fronts and the (relatively low) state of the interdisciplinary research, and the persistence dominance of both associations and research from the Global-North, but also, the comparative advantage and interest of each region. Further studies may consider other regions from the Global-South (i.e., Africa and other Asian Countries), additional technological innovation-related fields (i.e., STEM), complementary bibliographic databases/search engines (i.e., Google Scholar, Dimensions, Microsoft Academic), and non-redundant bibliometrics techniques (e.g., bibliographic coupling; main path analysis; more detailed network's structure measurements, among others).

**Dataset**

The dataset is available at the following permanent link: https://doi.org/10.34848/FK2/XHKWLL

**Acknowledgments**

The author would like to thank the Fudan-Latin America University Consortium (FLAUC), Fudan Development Institute, and Fudan University for providing the funding and support for conducting this research. Thanks to the Universidad del Rosario's School of Management and Business. Finally, thanks to Geoff Whyte, MBA, from Edanz Group (www.edanzediting.com/ac) for editing a draft of this manuscript.

**References**

Acedo, F. J., Barroso, C., Casanueva, C., & Galán, J. L. (2006). Co-Authorship in Management and Organizational Studies: An Empirical and Network Analysis*. *Journal of Management Studies*, *43*(5), 957–983. https://doi.org/10.1111/j.1467-6486.2006.00625.x

Amara, N., & Landry, R. (2012). Counting citations in the field of business and management: Why use Google Scholar rather than the Web of Science. *Scientometrics*, *93*(3), 553–581. https://doi.org/10.1007/s11192-012-0729-2

Anser, M. K., Zhang, Z., & Kanwal, L. (2018). Moderating effect of innovation on corporate social responsibility and firm performance in realm of sustainable development. *Corporate Social Responsibility and Environmental Management*, *25*(5), 799–806. https://doi.org/10.1002/csr.1495

Archambault, É., Campbell, D., Gingras, Y., & Larivière, V. (2009). Comparing bibliometric statistics obtained from the Web of Science and Scopus. *Journal of the American Society for Information Science and Technology*, *60*(7), 1320–1326. https://doi.org/10.1002/asi.21062

Australian Government - Department of Education. (n.d.). Ranking China's universities. Retrieved July 29, 2019, from Ranking China's universities website: https://internationaleducation.gov.au/News/Latest-News/Pages/Article-Ranking-Chinas-universities.aspx

Beall, J. (2015). *Beall's List: Potential, Possible, or Probable Predatory Scholarly Open-Access Publishers*.

Bergstrom, C. T., West, J. D., & Wiseman, M. A. (2008). The Eigenfactor$^{TM}$ Metrics. *Journal of Neuroscience*, *28*(45), 11433–11434. https://doi.org/10.1523/JNEUROSCI.0003-08.2008

Birkinshaw, J., Hamel, G., & Mol, M. J. (2008). Management innovation. *Academy of Management Review*, *33*, 825–845. https://doi.org/10.5465/amr.2008.34421969.

Bonilla, S. H., Almeida, C. M. V. B., Giannetti, B. F., & Huisingh, D. (2010). The roles of cleaner production in the sustainable development of modern societies: An introduction to this special issue. *Journal of Cleaner Production*, *18*(1), 1–5. https://doi.org/10.1016/j.jclepro.2009.09.001

Bornmann, L., Bowman, B., Bauer, J., Marx, W., Schier, H., & Palzenberger, M. (2014). Bibliometric standards for evaluating research institutes in the natural sciences. In *Beyond Bibliometrics: Harnessing Multidimensional Indicators of Scholarly Impact* (Blaise Cronin; Cassidy Sugimoto, pp. 201–224). Cambridge, USA: MIT Press.

Casper, S., & Murray, F. (2005). Careers and clusters: Analyzing the career network dynamic of biotechnology clusters. *J Eng Technol Manage JET M*, *22*(1–2), 51–74. https://doi.org/10.1016/j.jengtecman.2004.11.009

Ceretta, G. F., Dos Reis, D. R., & Da Rocha, A. C. (2016). Innovation and business models: A bibliometric study of scientific production on Web of Science database. *Gestao e Producao*, *23*(2), 433–444. https://doi.org/10.1590/0104-530X1461-14

Cetindamar, D., Wasti, S. N., Ansal, H., & Beyhan, B. (2009). Does technology management research diverge or converge in developing and developed countries? *Technovation*, *29*(1), 45–58. https://doi.org/10.1016/j.technovation.2008.04.002

Choi, D. G., Lee, Y.-B., Jung, M.-J., & Lee, H. (2012). National characteristics and competitiveness in MOT research: A comparative analysis of ten specialty journals, 20002009. *Technovation*, *32*(1), 9–18. https://doi.org/10.1016/j.technovation.2011.09.004

Clarivate Analytics. (2017). *Web of Science Fact Book*. Retrieved from Clarivate website: https://clarivate.com/wp-content/uploads/2017/05/d6b7faae-3cc2-4186-8985-a6ecc8cce1ee_Crv_WoS_Upsell_Factbook_A4_FA_LR_edits.pdf

Clarivate Analytics. (2018). WoS. Retrieved from Search website: http://apps.webofknowledge.com/WOS_GeneralSearch_input.do?product=WOS&search_mode=GeneralSearch&SID=6AMHraWex5T7wYhRXTx&preferencesSaved=

Confraria, H., Mira Godinho, M., & Wang, L. (2017). Determinants of citation impact: A comparative analysis of the Global South versus the Global North. *Research Policy*, *46*(1), 265–279. https://doi.org/10.1016/j.respol.2016.11.004

Cortés-Sánchez, J. D. (2018). A bibliometric outlook of the most cited documents in business, management and accounting in Ibero-America. *European Research on Management and Business Economics*, *26(1)*, 1-8. https://doi.org/10.1016/j.iedeen.2019.12.003

Cortés-Sánchez, J. D. (2019). Innovation in Latin America through the lens of bibliometrics: Crammed and fading away. *Scientometrics, 121,* 869-895. https://doi.org/10.1007/s11192-019-03201-0

De Carvalho, G. D. G., Cruz, J. A. W., De Carvalho, H. G., Duclós, L. C., & De Fátima Stankowitz, R. (2017). Innovativeness measures: A bibliometric review and a classification proposal. *International Journal of Innovation Science*, *9*(1), 81–101. https://doi.org/10.1108/IJIS-10-2016-0038

de Paulo, A. F., Carvalho, L. C., Costa, M. T. G. V., Lopes, J. E. F., & Galina, S. V. R. (2017). Mapping Open Innovation: A Bibliometric Review to Compare Developed and Emerging Countries. *Global Business Review*, *18*(2), 291–307. https://doi.org/10.1177/0972150916668600

Eigenfactor.org. (2019). Eigenfactor: About. Retrieved July 19, 2019, from FAQ website: http://www.eigenfactor.org/methods.htm

Favaretto, J. E. R., & Francisco, E. R. (2017). Exploring the archive of RAE-Revista de Administração de Empresas (1961-2016) in the light of bibliometrics, text mining, social network and geoanalysis [Exploração do acervo da RAE-Revista de Administração de Empresas (de 1961 a 2016) à luz da bibliometria, text mining, rede social e geoanálise]. *RAE Revista de Administracao de Empresas*, *57*(4), 365–390. https://doi.org/10.1590/S0034-759020170407

Garfield, E. (2015). The Agony and the Ecstasy—The History and Meaning of the Journal Impact Factor. *TEM Journal*, *4*(3), 1–22.

Gavel, Y., & Iselid, L. (2008). Web of Science and Scopus: A journal title overlap study. *Online Information Review*, *32*(1), 8–21. https://doi.org/10.1108/14684520810865958

Godin, B. (2015). Models of innovation: Why models of innovation are models, or what work is being done in calling them models? *Social Studies of Science*, *45*(4), 570–596. https://doi.org/10.1177/0306312715596852

Guan, J., & He, Y. (2007). Patent-bibliometric analysis on the Chinese science—Technology linkages. *Scientometrics*, *72*(3), 403–425. https://doi.org/10.1007/s11192-007-1741-1

Guan, J., & Ma, N. (2007). China's emerging presence in nanoscience and nanotechnology. A comparative bibliometric study of several nanoscience "giants." *Research Policy*, *36*(6), 880–886. https://doi.org/10.1016/j.respol.2007.02.004

Harzing, A.-W., & Giroud, A. (2014). The competitive advantage of nations: An application to academia. *Journal of Informetrics*, *8*(1), 29–42. https://doi.org/10.1016/j.joi.2013.10.007

Hirsch, J. E. (2005). An index to quantify an individual's scientific research output. *Proceedings of the National Academy of Sciences of the United States of America*, *102*(46), 16569–16572. https://doi.org/10.1073/pnas.0507655102

Huang, F. (2018). Quality deficit belies the hype. *Nature*, *564*, S70–S71. https://doi.org/10.1038/d41586-018-07694-2

Huang, L., Zhang, Y., Guo, Y., Zhu, D., & Porter, A. L. (2014). Four dimensional Science and Technology planning: A new approach based on bibliometrics and technology roadmapping. *Technological Forecasting and Social Change*, *81*(1), 39–48. https://doi.org/10.1016/j.techfore.2012.09.010

Kennedy, D. (2003). Multiple Authors, Multiple Problems. *Science*, *301*(5634), 733–733. https://doi.org/10.1126/science.301.5634.733

Kostoff, R. N. (2012). China/USA nanotechnology research output comparison-2011 update. *Technological Forecasting and Social Change*, *79*(5), 986–990. https://doi.org/10.1016/j.techfore.2012.01.007

Larivière, V., Haustein, S., & Mongeon, P. (2015). The Oligopoly of Academic Publishers in the Digital Era. *PLOS ONE*, *10*(6), e0127502. https://doi.org/10.1371/journal.pone.0127502

Lazzarotti, F., Dalfovo, M. S., & Hoffmann, V. E. (2011). A bibliometric study of innovation based on schumpeter. *Journal of Technology Management and Innovation*, *6*(4), 121–135. https://doi.org/10.4067/S0718-27242011000400010

Leonard-Barton, D. (1992). Core capabilities and core rigidities: A paradox in managing new product development. *Strategic Management Journal*, *13*(1 S), 111–125. https://doi.org/10.1002/smj.4250131009

Li, N., Chen, K., & Kou, M. (2017). Technology foresight in China: Academic studies, governmental practices and policy applications. *Technological Forecasting and Social Change*, *119*, 246–255. https://doi.org/10.1016/j.techfore.2016.08.010

Li, X., Zhou, Y., Xue, L., & Huang, L. (2015). Integrating bibliometrics and roadmapping methods: A case of dye-sensitized solar cell technology-based industry in China. *Technological Forecasting and Social Change*, *97*, 205–222. https://doi.org/10.1016/j.techfore.2014.05.007

Lopes, A. P. V. B. V., & De Carvalho, M. M. (2012). The evolution of the literature on innovation in cooperative relationships: A bibliometric study for the last two decades. *Gestao e Producao*, *19*(1), 203–217. Retrieved from Scopus.

Manjarrez, C. C. A., Pico, J. A. C., & Díaz, P. A. (2016). Industry Interactions in Innovation Systems: A Bibliometric Study. *Latin American Business Review*, *17*(3), 207–222. https://doi.org/10.1080/10978526.2016.1209036

Mendes, L., Amorim-Borher, B., & Lage, C. (2013). Patent applications on representative sectors of biotechnology in Brazil: An analysis of the last decade. *Journal of Technology Management and Innovation*, *8*(4), 91–102. Retrieved from Scopus.

Merton, R. K. (1968). The matthew effect in science. *Science*, *159*(3810), 56–62. Retrieved from Scopus.

Mingers, J., & Lipitakis, E. A. E. C. G. (2010). Counting the citations: A comparison of Web of Science and Google Scholar in the field of business and management. *Scientometrics*, *85*(2), 613–625. https://doi.org/10.1007/s11192-010-0270-0

Mongeon, P., & Paul-Hus, A. (2016). The journal coverage of Web of Science and Scopus: A comparative analysis. *Scientometrics*, *106*(1), 213–228. https://doi.org/10.1007/s11192-015-1765-5

OEC. (2018). OEC - China (CHN) Exports, Imports, and Trade Partners. Retrieved August 30, 2019, from https://oec.world/en/profile/country/chn/

Ortiz-Ospina, E., Beltekian, D., & Roser, M. (2018). Trade and Globalization—Our World in Data. Retrieved August 16, 2019, from https://ourworldindata.org/trade-and-globalization

Padilla-Ospina, A. M., Medina-Vásquez, J. E., & Rivera-Godoy, J. A. (2018). Financing innovation: A bibliometric analysis of the field. *Journal of Business and Finance Librarianship*, *23*(1), 63–102. https://doi.org/10.1080/08963568.2018.1448678

Peng, L., & Lai, L. (2014). A service innovation evaluation framework for tourism e-commerce in China based on BP neural network. *Electronic Markets*, *24*(1), 37–46. https://doi.org/10.1007/s12525-013-0148-0

Pilkington, A., & Teichert, T. (2006). Management of technology: Themes, concepts and relationships. *Technovation*, *26*(3), 288–299. https://doi.org/10.1016/j.technovation.2005.01.009

Qi, G. Y., Shen, L. Y., Zeng, S. X., & Jorge, O. J. (2010). The drivers for contractors' green innovation: An industry perspective. *Journal of Cleaner Production*, *18*(14), 1358–1365. https://doi.org/10.1016/j.jclepro.2010.04.017

Rosenbusch, N., Brinckmann, J., & Bausch, A. (2011). Is innovation always beneficial? A meta-analysis of the relationship between innovation and performance in SMEs. *Journal of Business Venturing*, *26*(4), 441–457. https://doi.org/10.1016/j.jbusvent.2009.12.002

SCImago. (2018). SCImago journal ranking. Retrieved from SCImago journal ranking website: https://www.scimagojr.com/

Scopus. (2018). Search. Retrieved from https://www.scopus.com/

Scopus. (2019). Scopus. Retrieved July 16, 2019, from How Scopus works website: https://www.elsevier.com/solutions/scopus/how-scopus-works

Silveira, F. F., & Zilber, S. N. (2017). Is social innovation about innovation? A bibliometric study identifying the main authors, citations and co-citations over 20 years. *International Journal of Entrepreneurship and Innovation Management*, *21*(6), 459–484.

Small, H., Sweeney, E., & Greenlee, E. (1985). Clustering the science citation index using co-citations. II. Mapping science. *Scientometrics*, *8*(5), 321–340. https://doi.org/10.1007/BF02018057

Small, Henry. (1973). Co-citation in the scientific literature: A new measure of the relationship between two documents. *Journal of the American Society for Information Science*, *24*(4), 265–269. https://doi.org/10.1002/asi.4630240406

Statista. (2018). China: GDP breakdown by sector. Retrieved August 30, 2019, from https://www.statista.com/statistics/270325/distribution-of-gross-domestic-product-gdp-across-economic-sectors-in-china/

Stilgoe, J., Owen, R., & Macnaghten, P. (2013). Developing a framework for responsible innovation. *Research Policy*, *42*(9), 1568–1580. https://doi.org/10.1016/j.respol.2013.05.008

Tanco, M., Escuder, M., Heckmann, G., Jurburg, D., & Velazquez, J. (2018). Supply chain management in Latin America: Current research and future directions. *Supply Chain Management*, *23*(5), 412–430. https://doi.org/10.1108/SCM-07-2017-0236

Teece, D. J. (2007). Explicating dynamic capabilities: The nature and microfoundations of (sustainable) enterprise performance. *Strategic Management Journal*, *28*(13), 1319–1350. https://doi.org/10.1002/smj.640

Teece, D. J., Pisano, G., & Shuen, A. (1997). Dynamic capabilities and strategic management. *Strategic Management Journal*, *18*(7), 509–533. https://doi.org/10.1002/(SICI)1097-0266(199708)18:7<509::AID-SMJ882>3.0.CO;2-Z

Tijssen, R. J. W. (2009). Internationalisation of pharmaceutical R&D: How globalised are Europe's largest multinational companies? *Technology Analysis and Strategic Management*, *21*(7), 859–879. https://doi.org/10.1080/09537320903182330

Tollefson, J. (2018). China declared world's largest producer of scientific articles. *Nature*, *553*, 390–390. https://doi.org/10.1038/d41586-018-00927-4

UN. (2018). Goal 9 -Build resilient infrastructure, promote inclusive and sustainable industrialization and foster innovation. Retrieved August 15, 2019, from https://sustainabledevelopment.un.org/sdg9

UNEP-WCMC. (2016). *El estado de la biodiversidad en América Latina y El Caribe*. Nairobi, Kenia.

United Nations. (2019). *Special edition: Progress towards the Sustainable Development Goals*. Retrieved from United Nations website: https://undocs.org/pdf?symbol=en/E/2019/68

van Eck, N. J., & Waltman, L. (2010). Software survey: VOSviewer, a computer program for bibliometric mapping. *Scientometrics*, *84*(2), 523–538. https://doi.org/10.1007/s11192-009-0146-3

Villasana, M. (2012). University-industry interactions in biotechnology: Implications for the development of a high-tech cluster. *International Journal of Learning and Intellectual Capital*, *9*(4), 429–447. https://doi.org/10.1504/IJLIC.2012.049618

Wuchty, S., Jones, B. F., & Uzzi, B. (2007). The increasing dominance of teams in production of knowledge. *Science*, *316*(5827), 1036–1039. https://doi.org/10.1126/science.1136099

Zupic, I., & Čater, T. (2015). Bibliometric Methods in Management and Organization. *Organizational Research Methods*, *18*(3), 429–472. https://doi.org/10.1177/1094428114562629

## Tables and figures

**Table 1 Query items for each bibliographic database**

| Bibliographic database | Keyword(s) | Subject area/Field | Subject categories | Source type | Author(s) affiliation(s) | Year | Results |
|---|---|---|---|---|---|---|---|
| WoS | Innovation; Innovación; innovação | Management; Business | Marketing and advertising; Forecasting; Planning; Administration; Organizational studies; Compensation; Strategy; Retailing; Consumer research, and management; Business history; Business ethics; Management science; Strategic planning and decision- | Article | China; LATAM | 2001-2018 | China: 2,336; LATAM: 1,272; Total (excluding repeated |

| | | making methods; Leadership studies; Total quality management. | | | | articles): 3,609 |
|---|---|---|---|---|---|---|
| Scopus | Business, management and accounting<br><br>Decision sciences | Accounting; Business and International Management; Business, Management and Accounting; Industrial Relations; Management Information Systems; Management of Technology and Innovation; Marketing; Organizational Behavior and Human Resource Management; Strategy and Management; Tourism, Leisure and Hospitality Management; Information Systems and Management; Management Science and Operations Research; Statistics, Probability and Uncertainty | Article | China; LATAM | 1996-2018 | China: 1,380; LATAM: 1,031; Total (excluding repeated articles): 2,411 |

Source: the author based on Scopus (2018) and WoS (2018). Note: the 20 Spanish-Portuguese speaking LATAM countries considered for the query, were: Argentina; Bolivia; Brazil; Chile; Colombia; Costa Rica; Cuba; Dominican Republic; Ecuador; El Salvador; Guatemala; Honduras; Mexico; Nicaragua; Panama; Paraguay; Peru; Puerto Rico; Uruguay; and Venezuela.

**Table 2 Descriptive statistics of WoS and Scopus samples**

| Region | | WoS | | | | Scopus | | |
|---|---|---|---|---|---|---|---|---|
| | | # Authors | Impact Factor | Eigenfactor | Citations | # Authors | Journal H Index | Citations |
| China | N | | 2,336 | | | | 1,377 | |
| | Open access | | 171 (7%) | | | | 80 (5%) | |
| | Uncited | - | - | - | 527 (22%) | - | - | 201 (14%) |
| | Mean | 2.94 | 2.17 | 0.005 | 20.3 | 3.02 | 64.53 | 17.19 |
| | Median | 3 | 1.83 | 0.002 | 4 | 3 | 56 | 5 |
| | Min | 1 | 0 | 0 | 0 | 1 | 0 | 0 |
| | Max | 8 | 9.28 | 0.05 | 872 | 9 | 232 | 679 |
| | SD | 1.12 | 1.68 | 0.008 | 43.94 | 1.198 | 50.35 | 33.14 |
| LATAM | N | | 1,272 | | | | 1,023 | |
| | Open access | | 655 (51%) | | | | 230 (22%) | |
| | Uncited | - | - | - | 801 (62%) | - | - | 221 (21%) |
| | Mean | 2.91 | 1.00 | 0.002 | 18.2 | 2.97 | 42.9 | 12.9 |
| | Median | 3 | .00 | 0 | 7 | 3 | 18 | 3 |
| | Min | 1 | 0 | 0 | 2 | 1 | 0 | 0 |
| | Max | 11 | 6.70 | 0.05 | 371 | 28 | 209 | 469 |
| | SD | 1.30 | 1.47 | 0.005 | 33.92 | 1.545 | 45.57 | 28.24 |
| Total | N | | 3,608 | | | | 2,360 | |
| | Open access | | 826 (22%) | | | | 310 (13%) | |
| | Sum | - | - | - | 45,357 | - | - | 30,116 |
| | Uncited | - | - | - | 1,328 (36%) | - | - | 422 (17%) |
| | Mean | 2.93 | 1.76 | 0.00405 | 19.89 | 3 | 55.31 | 12.55 |
| | Median | 3 | 1.41 | 0.001 | 4 | 3 | 41 | 4 |
| | Min | 1 | 0 | 0 | 0 | 1 | 0 | 0 |
| | Max | 11 | 9.28 | 0.058 | 872 | 28 | 232 | 679 |
| | SD | 1.192 | 1.71 | 0.007945 | 42.43 | 1.357 | 49.53 | 31.22 |

Source: the author based on Scopus (2018) and WoS (2018). Note: the mean citation just considers articles with at least 1 citation.

**Table 3 ANOVAs and post hoc Tukey HSD tests summary comparing both regions for number of authors, citations, Journals' Impact Factor, Eigenfactor, and journals' H Index**

| | WoS | | | | | Scopus | | | | |
|---|---|---|---|---|---|---|---|---|---|---|
| Variable | Groups | ANOVA | p<.05 Y / N | | Post hoc Tukey HSD | Variable | Groups | ANOVA | p<.05 Y / N | Post hoc Tukey HSD |
| Number of authors | G1: China (n=2,336); G2: LATAM (n=1,272) | [F(1,3606)=.536, p=.464] | | X | | Number of authors | G1: China (n=1,377); G2: LATAM (n=1,023) | [F(1,2398)=.794, p=.373] | X | |
| Citations | | [F(1,3606)=46.801, p=.000] | X | | G1 ($\bar{x}$=20.3, $\sigma$=43.9); G2 ($\bar{x}$=18.2, $\sigma$=33.9) | Citations | | [F(1,2397)=12.030, p=.001] | X | G1 ($\bar{x}$=17.1, $\sigma$=33.1); G2 ($\bar{x}$=12.9, $\sigma$=28.2) |
| JIF | | [F(1,3606)=434.194, p=.000] | X | | G1 ($\bar{x}$=2.17, $\sigma$=1.6); G2 ($\bar{x}$=1.0, $\sigma$=1.4) | Journals' H Index | | [F(1,2805)=19.555, p=.000] | X | G1 ($\bar{x}$=64.5, $\sigma$=50.3); G2 ($\bar{x}$=42.9, $\sigma$=45.5) |
| Eigenfactor | | [F(1,3606)=109.331, p=.000] | X | | G1 ($\bar{x}$=0.005, $\sigma$=0.008); G2 ($\bar{x}$=0.002, $\sigma$=0.005) | Group of authors-Citations | G1: individual (n=223); G2: couples (n=679); G3: trios (n=751); G4: packs (n=747) | [F(3,2396)=.826, p=.48] | X | |
| Group of authors-Citations | G1: individual (n=340); G2: couples (n=1,027); G3: trios (n=1,202); G4: packs (four or more authors, n=1,039) | [F(3,3604)=1.587, p=.190] | | X | | Group of authors-Journals' H Index | | [F(3,2324)=13.00, p=.000] | X | G4 ($\bar{x}$=65.5, $\sigma$=49.76) & G1 ($\bar{x}$=49.4, $\sigma$=46.1) G2 ($\bar{x}$=50.5, $\sigma$=49.6) G3 ($\bar{x}$=56.6, $\sigma$=48.2) |

Source: the author based on WoS (2018) and Scopus (2018). SPSS was used for ANOVAs and post hoc tests

**Table 4 Top-ten cited papers in each search engine/bibliographic database**

| Scopus | | | | | | | | |
|---|---|---|---|---|---|---|---|---|
| Region | Citations | Authors | Lead author affiliation | Year | Title | Source | Source h-index |
| China | 679 | Jensen M.B., Johnson B., Lorenz E., Lundvall B.A., | Department of Marketing and Statistics, Aarhus School of Business, University of Aarhus, Denmark | 2007 | Forms of knowledge and modes of innovation | Research Policy | 191 |
| China | 403 | Zeng S.X., Xie X.M., Tam C.M., | Antai School of Management, Shanghai Jiaotong University, China | 2010 | Relationship between cooperation networks and innovation performance of SMEs | Technovation | 102 |
| China | 325 | Liu X., White S. | National Research Center of Science and Technology for Development, Research Center for Innovation Strategy and Management, China | 2001 | Comparing innovation systems: A framework and application to China's transitional context | Research Policy | 191 |
| China | 262 | Guan J., Ma N. | School of Management, Beijing Univ. of Aero./Astronautics, China | 2003 | Innovative capability and export performance of Chinese firms | Technovation | 102 |
| China | 224 | Xiao J.Z., Yang H., Chow C.W. | Cardiff University, United Kingdom | 2004 | The determinants and characteristics of voluntary Internet-based disclosures by listed Chinese companies | Journal of Accounting and Public Policy | 58 |
| LATAM | 469 | Stilgoe J., Owen R., Macnaghten P., | University of Exeter Business School, University College London, United Kingdom | 2013 | Developing a framework for responsible innovation | Research Policy | 191 |
| LATAM | 394 | Perez C. | Department of Technological Development, Ministry of Industry, Venezuela | 1983 | Structural change and assimilation of new technologies in the economic and social systems | Futures | 66 |
| LATAM | 264 | Bessant J., Caffyn S. | Centre for Research in Innovation Management, University of Brighton, United Kingdom | 1997 | High-involvement innovation through continuous improvement | International Journal of Technology Management | 48 |
| LATAM | 189 | Naranjo-Valencia J.C., Jiménez-Jiménez D., Sanz-Valle R., | Facultad de Administración, Universidad Nacional de Colombia, Colombia | 2011 | Innovation or imitation? The role of organizational culture | Management Decision | 77 |
| LATAM | 179 | Ramdani B., Kawalek P., Lorenzo O. | Manchester Business School, The University of Manchester, United Kingdom | 2009 | Predicting SMEs' adoption of enterprise systems | Journal of Enterprise Information Management | 48 |

| | | | | WoS | | | |
|---|---|---|---|---|---|---|---|
| China | 872 | Zhu Q.H., Sarkis, J. | School of Management, Dalian University of Technology, China | 2004 | Relationships between operational practices and performance among early adopters of green supply chain management practices in Chinese manufacturing enterprises | Journal Of Operations Management | 166 |
| China | 593 | Zhou K.Z., Yim C.K., Tse D.K. | Department of Marketing, School of Business, University of Hong Kong, Hong Kong | 2005 | The effects of strategic orientations on technology- and market-based breakthrough innovations | Journal Of Marketing | 218 |
| China | 588 | Li H.Y., Atuahene-Gima K. | Texas A&M University | 2001 | Product innovation strategy and the performance of new technology ventures in China | Academy Of Management Journal | 283 |
| China | 552 | Atuahene-Gima K. | Department of Marketing and Innovation Management, China Europe International Business School (CEIBS), China | 2005 | Resolving the capability-rigidity paradox in new product innovation | Journal Of Marketing | 218 |
| China | 486 | Chau P.Y.K., Hu P.J.H. | School of Business, Faculty of Business and Economics, University of Hong Kong, Hong Kong | 2001 | Information technology acceptance by individual professionals: A model comparison approach | Decision Sciences | 97 |
| LATAM | 371 | Stilgoe, J., Owen R., Macnaghten P. | University of Exeter Business School, University College London, United Kingdom | 2013 | Developing a framework for responsible innovation | Research Policy | 191 |
| LATAM | 293 | Owen R., Macnaghten P., Stilgoe J. | University of Exeter Business School, United Kingdom | 2012 | Responsible research and innovation: From science in society to science for society, with society | Science and Public Policy | 51 |
| LATAM | 244 | Sosna M., Trevinyo-Rodriguez R., Velamuri S. | IESE Business School, Spain | 2010 | Business Model Innovation through Trial-and-Error Learning The Naturhouse Case | Long Range Planning | 89 |
| LATAM | 206 | Vassolo, R.S., Anand J., Folta T.B. | Universidad Austral, Argentina | 2004 | Non-additivity in portfolios of exploration activities: A real options-based analysis of equity alliances in biotechnology | Strategic Management Journal | 232 |
| LATAM | 187 | Yam R.C.M., Guan J.C., Pun K.F., Tang E.P.Y. | Dept. of Manufacturing Engineering, City University of Hong Kong, Hong Kong | 2004 | An audit of technological innovation capabilities in Chinese firms: some empirical findings in Beijing, China | Research Policy | 191 |

Source: the author based on Scopus (2018) and WoS (2018).

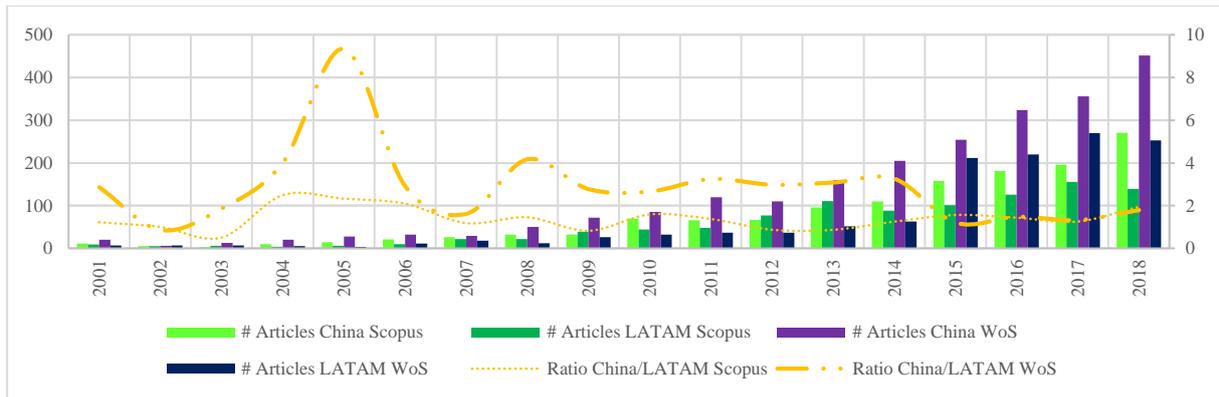

**Figure 1** China and LATAM output and ratio for both WoS and Scopus 2001-2018. Source: the author based on WoS (2018) and Scopus (2018)

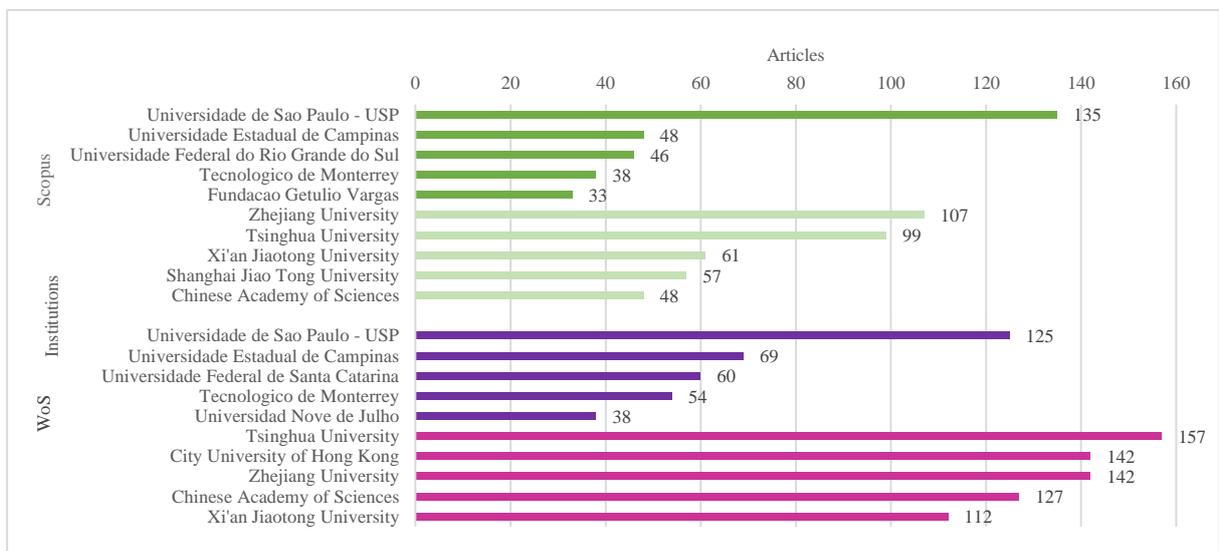

**Figure 2** China and LATAM output by institution for both WoS and Scopus 2001-2018. Source: the author based on WoS (2018) and Scopus (2018)

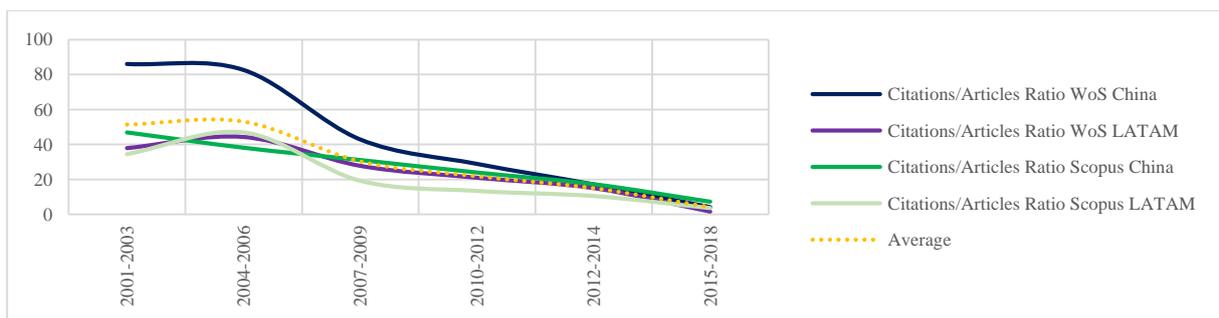

**Figure 3** China and LATAM citation/articles ratio for both WoS and Scopus 2001-2018. Source: the author based on WoS (2018) and Scopus (2018)

a) **Co-occurrence network, China**

b) **Co-occurrence network, LATAM**

**Figure 4 Co-occurrence network of China and LATAM. Source: the author based on Scopus (2018) using VOSviewer**

a) Co-occurrence network, China

**b) Co-occurrence network, LATAM**

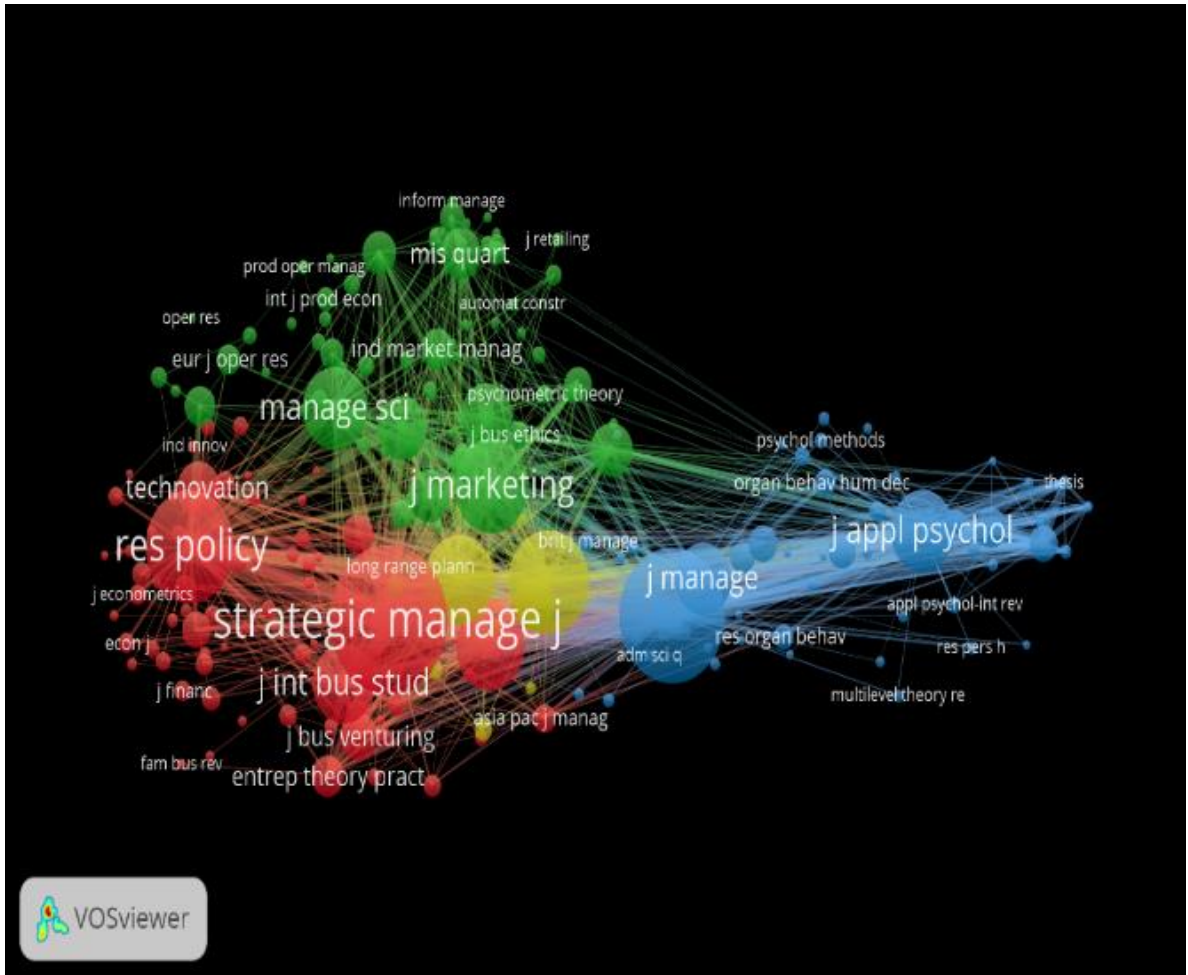

**c) Co-citation network, China**

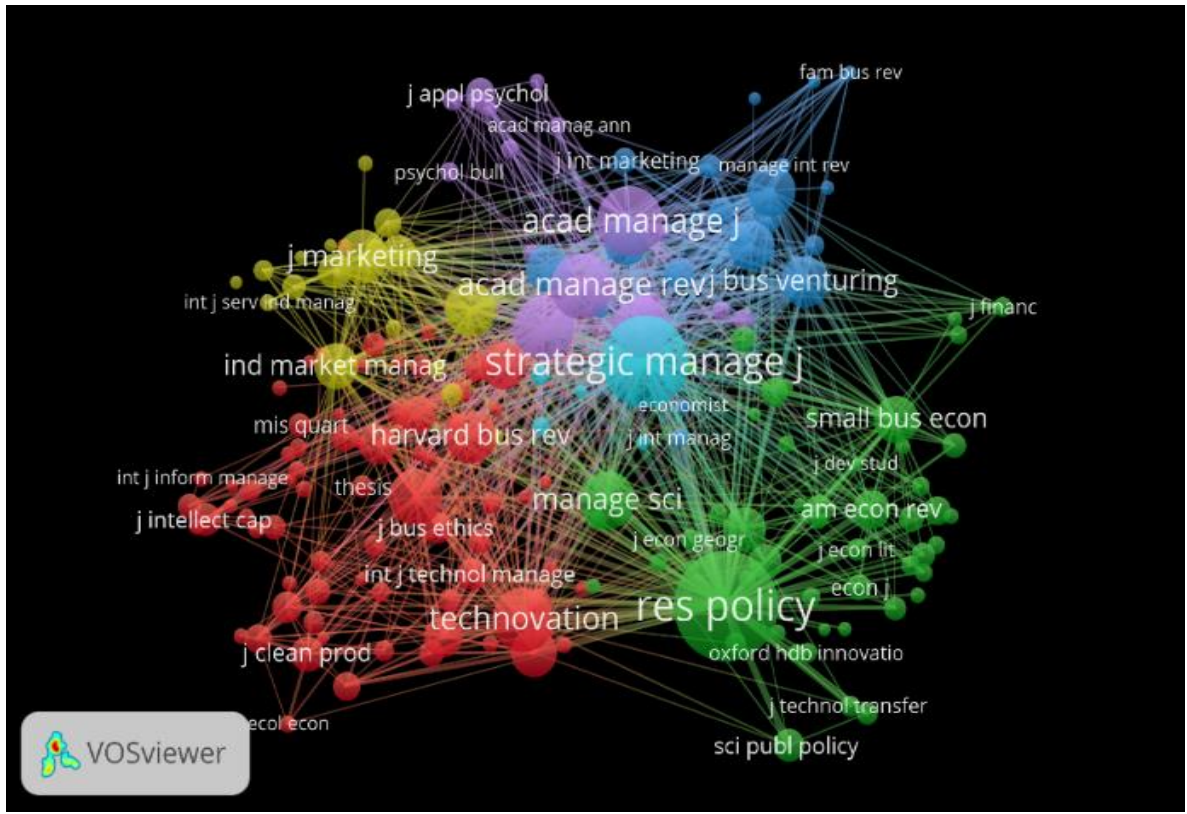

**d) Co-citation network, LATAM**

**Figure 5 Co-occurrence and co-citation networks of China and LATAM. Source: the author based on WoS (2018) using VOSviewer**